\documentclass[aps,prl,reprint,superscriptaddress]{revtex4-2}

\usepackage{graphicx}
\usepackage{hyperref}%
\usepackage{siunitx}    
\usepackage{amsmath,amssymb,bm,mathtools}
\usepackage{physics}
\usepackage{bbm}
\usepackage{xcolor}
\usepackage{comment}
\usepackage{booktabs} 
\usepackage{multirow}

\newcommand{\srb}{$^{88}$Sr~}

\newcommand{\srsc}{$^1$S$_0$\,--\,$^3$P$_1$\,}
\newcommand{\srclock}{$^1$S$_0$\,--\,$^3$P$_0$\,}

\begin{document}


\title{Measurement and control of {the} interaction frequency shift in bosonic optical lattice clocks}


\author{J. P. Salvatierra}
\affiliation{Istituto Nazionale di Ricerca Metrologica, Strada delle Cacce 91, 10135 Torino, Italy}
\author{M. Barbiero}
\affiliation{Istituto Nazionale di Ricerca Metrologica, Strada delle Cacce 91, 10135 Torino, Italy}
\author{G. Bertaina}
\affiliation{Istituto Nazionale di Ricerca Metrologica, Strada delle Cacce 91, 10135 Torino, Italy}
\author{D. Calonico}
\author{F. Levi}
\author{M. G. Tarallo}
\email[]{m.tarallo@inrim}
\affiliation{Istituto Nazionale di Ricerca Metrologica, Strada delle Cacce 91, 10135 Torino, Italy}


\date{\today}

\begin{abstract}    

We report precise measurements of inter-level interactions in a bosonic optical lattice clock based on \srb atoms. We observe a nonlinear density dependence of the clock shift, even without reaching quantum degeneracy. In a 2D lattice, the Rabi line shape exhibits an interaction sideband consistent with a collective spin model, while in a 1D lattice the shift is modified by density-induced dephasing. These findings, combined with a careful choice of interrogation detuning and atomic density, can enable operation at a net-zero systematic density shift in \srb lattice clocks. We discuss the implications of these findings in many-body physics, quantum simulation, and precision isotope shift measurements, which provide a powerful probe for new physics beyond the Standard Model.
\end{abstract}


\maketitle

\paragraph{Introduction}

Optical lattice clocks (OLCs) are currently at the forefront of precision metrology, offering the potential for unprecedented accuracy in timekeeping~\cite{Ludlow2015,Derevianko_2011}. Their remarkable achievements include pushing the boundaries of measurement precision to the $10^{-18}$ level and beyond~\cite{McGrew_2018,Bothwell2022,Aeppli2024}, enabling new tests of fundamental physics such as searches for dark matter~\cite{Filzinger2023}, investigations into the potential variation of fundamental constants~\cite{Safronova2019,Sherrill2023}, and relativistic geodesy~\cite{Shinkai2025}. These advancements are also pivotal in contributing to the future redefinition of the SI second, moving towards an optical standard~\cite{RIEHLE2015}. However, significant challenges remain in their development and deployment. Precise mitigation of systematic shifts arising from atom-atom interactions has been theoretically modeled~\cite{Rey_2014} and experimentally achieved in fermionic OLCs~\cite{Aeppli_2022}, while strong s-wave interactions in bosonic species lead to significant density-dependent shifts and decoherence, thereby preventing bosonic clocks from competing with their fermionic counterparts in terms of ultimate accuracy, despite their simpler atomic structure and potentially longer coherence times. However, precise spectroscopy of clock transitions in bosonic isotopes has {recently gained} interest because of the possibility of probing new long-range interactions beyond the Standard Model by isotope shift spectroscopy~\cite{Berengut_2018,Ono2022,Berengut_2025}.

To overcome the s-wave interaction limit in bosonic OLCs, several technical expedients have been utilized, such as eliminating multiply occupied lattice sites by photoassociation~\cite{Zelevinsky_2006}, or employing 3D optical lattices with very low average occupation number~\cite{Akatsuka_2008}, or both. This results in other systematic shifts, such as those arising from imperfect lattice potentials and their polarization control. Previous experimental measurements of the atom-atom interaction shift on bosonic OLCs, and in particular of the \srb OLC, have considered only the linear dependence with respect to the number of atoms (or the atomic density)~\cite{Lisdat2009,Takano_2017,Origlia_2018}. {However, simple mean-field elastic interaction theory~\cite{Killian_1998,Harber_2002} of the linear density shift} was unable to correctly reproduce the values of the elastic s-wave scattering lengths of the \srb clock states~\cite{Winfred2010}.

The possibility of finding a {nonlinear} regime in which the effect of interactions can be suppressed or reduced by coherent manipulation of the atomic ensemble considered as a pseudo-spin ensemble has been extensively studied in the framework of atom interferometry~\cite{Bonnin_2019}, mainly in terms of increased atomic coherence.

In this work, we {probe the interaction properties of the \srb lattice clock by precise density shift measurements, also enhancing the interaction in a two-dimensional lattice. We} show an optical lattice clock based on bosonic atoms exhibiting a nonlinear clock frequency shift as a function of the number of atoms in each well of the lattice, with a nonlinearity amplified or suppressed by tuning the clock-locking point on the Rabi resonance. The experimental results are analyzed both in terms of unitary many-body and dissipative mean-field theory. The density shift nonlinearity allows the \srb optical lattice clock to be tuned to a ``magic'' density at which the collisional shift is canceled.

\begin{figure*}[tb]
    \centering
    \includegraphics[width=0.995\linewidth]{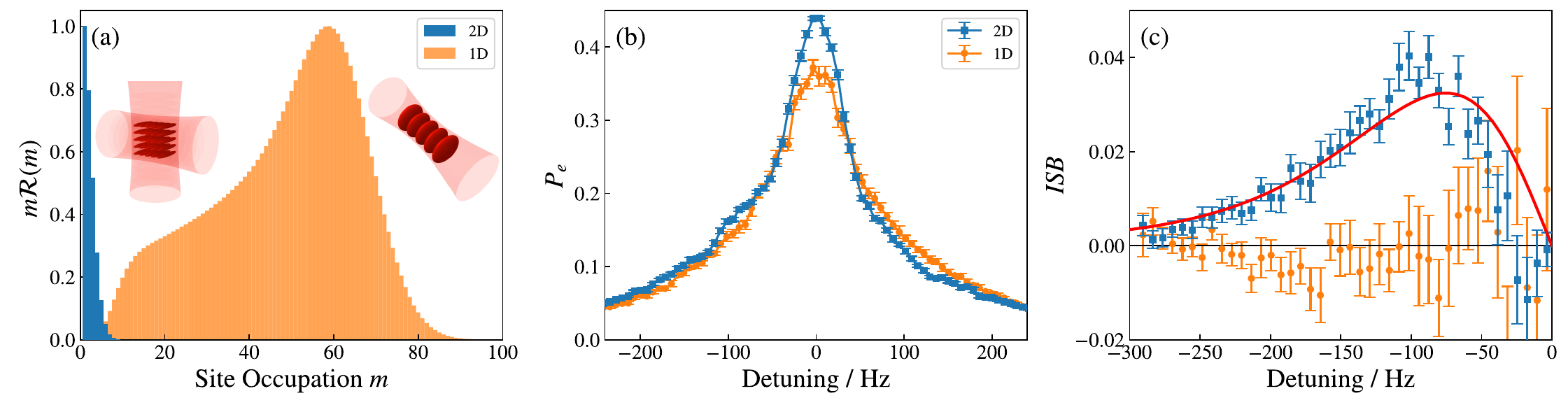}
    \caption{Rabi spectroscopy of interacting bosons in optical lattices. a) Fraction of the atomic population contributing to the spectroscopic signal with a given site occupancy $m$ for the 1D and 2D lattice configurations for $N_\text{tot} =  7\times10^4$, with a pictorial view of their confinement volumes.
    b)  Measured Rabi line shapes of the excited fraction for $g$ to $e$ interrogation for $N_\text{tot} = 7\times10^4$ atoms, taken in a 1D lattice ({orange circles}) and in a 2D lattice (blue squares). 
    While both profiles show a substantial broadening, the 2D profile is clearly asymmetric, revealing a many-body interaction structure. c) The reconstructed resolved atomic interaction sideband is plotted for both lattice topologies. The data from each line shape are collected into bins 3 Hz wide.  
    }
    \label{fig:Rabi-spec}
\end{figure*}

\paragraph{Theory} In essence, the optical lattice clock physical system~\cite{Derevianko_2011} consists of a series of independent atomic ensembles tightly confined in the maxima or minima, depending on the polarizability of the atom at the trapping frequency, of a periodic potential generated by interfering far-off resonance intense Gaussian laser beams. The number of standing waves employed in the system determines the dimensionality of the resulting lattice~\cite{Bloch2005}. Tuning the trapping laser frequency to cancel the AC Stark shift~\cite{Katori1999}, the tight confinement along the lattice direction suppresses the motional systematic effects~\cite{Dicke1953}, while thermal motion in the dimensions transverse to the lattice direction allows the atoms to interact, both elastically and inelastically.

Assuming operation of an optical lattice with negligible tunneling between nearest-neighbor lattice wells, {as well as other typical conditions like longitudinal freezing, stationary thermal radial state at temperature $T_r$, weak backaction of spin dynamics on the radial distribution~\cite{Lhuillier_1982}, and short radial correlation time compared with the other clock times~\cite{Rey_2014}, it is possible to trace over the radial modes and keep only an effective spin dynamics, obtaining} the Hamiltonian governing the dynamics of an ensemble of $N$ two-level bosons in a single well, driven by an external clock field as the spin Hamiltonian~\cite{suppMaterial}:

\begin{equation}\label{eq:Hspin}
\hat{H} = -\hbar\delta\hat{S}_z - \hbar\Omega\hat{S}_x +  C(N-1)\hat{S}_z+\chi\hat{S}_z^2\,.    
\end{equation}
{H}ere $\hat{\vec{S}} =\{\hat{S}_x,\hat{S}_y,\hat{S}_z\}$ are the collective spin operators {derived from the bosonic field operators}, with $\hat{S}_z \equiv (\hat{N}_e-\hat{N}_g)/2$ the atomic inversion operator and $e$ and $g$ label the excited and ground clock levels, respectively, while $C = (U_{ee}-U_{gg})/2$ and $\chi = (U_{ee}+U_{gg}-2U_{eg})/2$ represent the linear and nonlinear interaction energy scales, respectively, as a function of the {thermally averaged over the radial modes two-body} interaction strengths $U_{ij}=\kappa_{ij}^{(2)}(4\pi\hbar^2a_{ij})/M\,\int|w(\mathbf{r})|^4d^3r$, where $a_{ij}$ are the scattering lengths for levels $i, j${, the $\kappa_{ij}^{(2)}$ coefficients encode same-mode pair correlations which for an ideal thermal bosonic radial mode is equal to 2}, and $w(\mathbf{r})$ are the {thermally averaged Wannier functions~\cite{suppMaterial}}. The latter integral can be seen as the single atom density {$n_0$} and depends on the details of the depth of the lattice confinement and the thermal energy of each atom~\cite{suppMaterial}. The effect of non-vanishing interaction energies on the optical clock experimental observable, the excitation probability $P_e =\langle \hat{S}_z\rangle/N+1/2$, is both a shift of the resonance peak and an asymmetric lineshape, affecting the operating clock frequency which obeys the simple relation
{
\begin{equation}\label{eq:lock}
P_e(\delta_{\rm int}+\delta_{\text{lock}})-P_e(\delta_{\rm int}-\delta_{\text{lock}}) =0\,,
\end{equation}
}
where $\delta_{\text{lock}}$ is the clock laser offset frequency used in the clock cycle to extrapolate the clock resonance frequency, and $\delta_{\text{int}}$ is the detuning of the clock frequency with respect to the unperturbed atom. 

The dynamics of the system can experience different regimes depending on the magnitude of the interaction parameters with respect to the Rabi frequency $\Omega$ {and the lattice site occupation value $m$. Regarding the interaction strengths, a}t typical temperatures ($\sim 1\,\mu$K) and lattice depths ($\sim 100\,E_r$) of a 1D optical lattice clock, we expect that $\hbar\Omega\gg \chi,C$, so that no effect on the Rabi spectrum should be visible. On the other hand, reducing the dimensionality of the system by introducing a second lattice can easily shift the system from the weakly to the strongly interacting regime even at $\mu$K temperatures and relatively low density~\cite{Swallows_2011,Swallows_2012}, with visible interaction sidebands.

In the case of high occupation number, \srb has previously exhibited a strong dephasing rate~\cite{Lisdat2009}, which may imply a leakage from the collective spin manifold due to local radial mode-changing interactions. These dissipative effects can be included at the microscopic level (see \cite{suppMaterial}), however the solution of such dissipative many-body model is cumbersome. Therefore, we efficiently simulate the dynamics of the expectation values of the spin operators reduced to c-numbers in the mean-field approximation~\cite{Band_Collisionalshiftsopticallattice_2006}, including dissipation. This is equivalent to approximating quadratic spin correlators as products of expectation values of collective spin operators. Introducing the single-particle Bloch vector variables {$u=2\langle \hat S_x\rangle/N$, $v=2\langle \hat S_y\rangle/N$, $w=2\langle \hat S_z\rangle/N$}, the nonlinear optical Bloch equations including {interaction-induced dephasing} are (see Suppl. Mat.~\cite{suppMaterial} for a microscopic derivation, including two-body losses)

\begin{align}
    \dot u&=(\delta-\delta\nu_{\rm int}(w))v-\frac{\gamma_{\rm deph}}{4}(N-1)\left(1-w\right)u,\nonumber\\
    \dot v&=\Omega w -(\delta-\delta\nu_{\rm int}(w))u-\frac{\gamma_{\rm deph}}{4}(N-1)\left(1-w\right)v,\nonumber\\
    \dot w&=-\Omega v.\label{eq:mfeqs}
\end{align}
where {$\hbar\delta\nu_{\rm int} = C(N-1)+\chi Nw$} is the nonlinear interaction shift, which in the limit of $N\gg1$ reduces to the well-known mean-field density shift formula~\cite{Harber_2002}, while $\gamma_{\rm deph}$ is the dephasing rate constant of the density dependent coherence relaxation. From $\delta\nu_{\rm int}$ one can appreciate that it is possible to tune $w$ (or $\langle\hat{S}_z\rangle$ from (\ref{eq:Hspin})), by means of $\delta_{\rm lock}$, to reduce or cancel out the interaction shift or, vice-versa, to find a {``magic''} lattice atomic number $N$ at which the interaction shift vanishes, depending on the interaction energies $U_{ij}$. Interaction shift cancellation in fermionic clocks in a Rabi spectroscopy protocol has already been devised~\cite{Lee_2016}, but without including the nonlinear interaction effect.

In order to take into account the nonlinear interaction term and thus accurately calculate the interaction frequency shifts, we numerically solve both the unitary many-body Schr\"odinger equation using Eq.~\eqref{eq:Hspin} and the dissipative optical Bloch equations~\eqref{eq:mfeqs}. {Finally, we} compare {their} predictions with our data {by integrating over lattice site occupation distribution $m\mathcal{R}(m)$, where {$\mathcal{R}(m)$} is the probability of having $m$ atoms at any lattice site which is obtained from a Poisson distribution, with the mean occupation number at each site determined by a Gaussian distribution and averaged over all sites~\cite{Rey_2014} (see Fig.~\ref{fig:Rabi-spec}(a))}. 

\paragraph{Experimental setup.} The experimental setup for the \srb optical lattice clock used in these experiments has been extensively described in previous works~\cite{Barbiero_2022,Barbiero_2023}. After laser cooling and trapping, the ultracold atomic ensemble is interrogated by the clock laser beam aligned along the horizontal lattice (namely, the $z$ axis). Additionally, a second optical lattice has been included along the vertical direction (the $y$ axis), orthogonal to the existing lattice. The two lattice laser beams are independently controlled by acousto-optic modulators and detuned from each other by roughly 160 MHz to avoid unwanted interference and phase instabilities. The geometrical and thermodynamic properties of the 1D and 2D lattice configurations are characterized by precise sideband spectroscopy, time-of-flight absorption imaging, and parametric heating measurements. Typical clock operations are performed with a lattice temperature of 1.2 $\mu$K and 2 $\mu$K for the 1D and 2D lattice, respectively, and trap frequencies of $\nu_z$ = 68 kHz and $\nu_y$ = 14 kHz.

By controlling the laser power of the cold atomic source~\cite{Barbiero2020} we change the atomic population loaded into the optical lattice. In the 1D lattice case, atoms are loaded from a MOT operating on the \srsc transition, which produces an approximately Gaussian spatial distribution with a standard deviation $\sigma_z^{\text{1D}} = \sigma_{\text{MOT}}=\SI{192(7)}{\micro m}$.

\begin{figure}[t]
    \centering
    \includegraphics[width=0.99\linewidth]{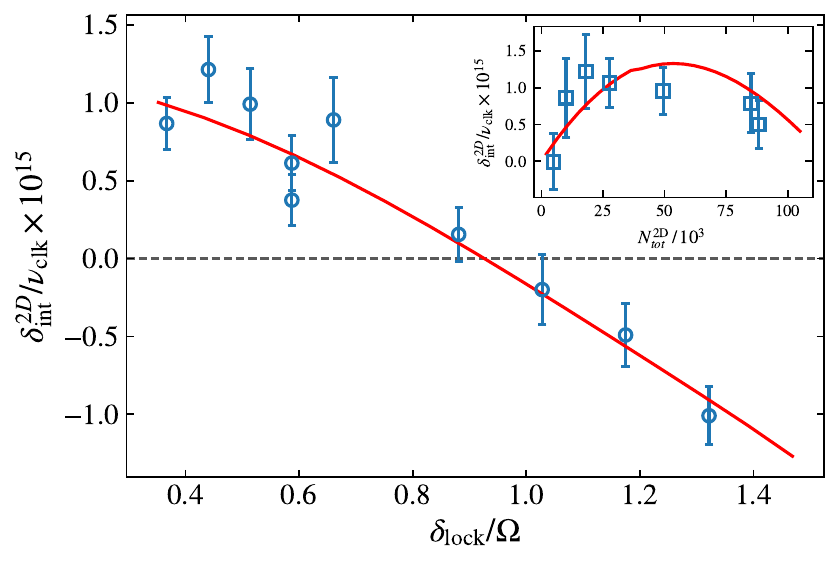}
    \caption{Observation of clock frequency shift in a 2D lattice as function of the excitation fraction $P_e$ as set by the locking point $\delta_{\text{lock}}$ {(main panel) fitted by spin model integration (solid red line), with the absolute shift at the reference point pinned by the density shift measurement at different $N_\text{tot}$ (in the inset)}. See main text for details.}
    \label{fig:2D-exc-frac}
\end{figure}

For the 2D lattice, to avoid loading lattice sites away from the intersection of the two orthogonal beams, the two lattice beams are {alternately switched} on and off adiabatically, so that the final spatial {extent} of the 2D lattice is due to the transverse width of the vertical lattice for the $z$ direction, and by the transverse width of the horizontal lattice in the $y$ axis. The resulting spatial widths are estimated to be {$\sigma_z^{\text{2D}} = 70(5)\,\mu$m and $\sigma_y^{\text{2D}} = 12(5)\,\mu$m}. We also apply to the 2D lattice case the Gaussian approximation for determining the lattice occupation probability in our numerical treatment. 

Since the interaction strength heavily depends on the single atom density, we make use of the measurement of the two-body inelastic scattering rate for the clock excited state, which has the same functional dependence on the atomic volume~\cite{Bouganne_2017}, as a check for both the 1D and 2D lattices~\cite{suppMaterial}. For the 1D lattice, the single atom density is about $1.2\times10^{10}$ cm$^{-3}$, while the two-body loss parameter {$\beta_{^3P_0} = 21(7)$ $\mu$m$^3$/s}, which is consistent with the previously measured value{~\cite{Traverso_2009,Dolde2025}}. For the 2D lattice, the single atom density is evaluated at about {\SI{7.9e11}{cm^{-3}}}.

Interleaved frequency measurements between different values of the atomic population allow us to precisely address the interaction shift. The two lattice populations are stable during the hour-long averaging time within 10\%. Rabi spectroscopy on the atomic ensemble is performed with a clock pulse of $T_\pi = 15$ ms, corresponding to a $\Omega=2\pi\times$\SI{34}{Hz}. The relative clock frequency instability for these measurements is about $1\times10^{-14}/\sqrt{\tau}$, mainly limited by the residual Dick effect due to the residual phase noise of the local oscillator frequency reference and the low duty cycle ($\sim 4$\%). This allows us to estimate frequency differences with a statistical uncertainty of about $3\times10^{-16}$ after \SI{1e3}{s} of integration.

\paragraph{Results}

We first consider experimental lineshapes for highly populated ($>10^4$) 1D and 2D lattices, as shown in Fig.~\ref{fig:Rabi-spec}. These lineshapes were obtained by scanning {the} laser frequency across the clock resonance several times and averaging the atomic response at each detuning. In the 1D lattice, we observe a nearly {symmetric peak whose broadening is mainly due to contributions from lattice sites with high occupation numbers $m$, as expected from the occupation distribution shown in Fig.~\ref{fig:Rabi-spec}(a)}. However, in the 2D lattice, strong interactions in multiply occupied lattice tubes lead to both a slight asymmetry at negative detuning and a slightly increased excitation fraction at resonance. 
This can also be understood as a consequence of the relative majority of singly populated lattice sites. 
An interaction sideband can be constructed by subtracting the excitation values at positive detunings to their corresponding negative values~\cite{Bishof11}, as reported in {Fig.~\ref{fig:Rabi-spec}(c)}. The negative detuning of the interaction sideband implies that the $U_{eg}-U_{gg}$ energy difference is negative and {due to} the very small value of $a_{gg}$~\cite{Stellmer2013} {(which we neglect in the following)}, $a_{eg}$ must also be negative. The interaction sideband can be heuristically modeled as the result of the sum of the single-atom excitation probabilities of the {$m$}-populated lattice tubes, i.e. {$\text{ISB}(\delta) \propto \sum_m m\mathcal{R}(m)[f(\delta-(m-1)U_{eg}/h,\Gamma)-f(-\delta-(m-1)U_{eg}/h,\Gamma)]$}, {where} $f(\delta,\Gamma)$ is the Rabi spectral response with a full width at half-maximum $\Gamma$ that could incorporate broadening effects. This model provides a basic understanding of the data, as shown by the line in Fig.~\ref{fig:Rabi-spec}{(c)}, where {$U_{eg}/h = $ -32(8)} Hz is the inter-level interaction energy.

We {also} directly measure the effect of interactions on the OLC frequency in the 2D lattice by interleaved self-comparison. The interaction-induced frequency shift at a fixed total atom number {($N_\text{tot}^0 =8.4\times10^3$)} {with respect to a reference lock point $\delta_{\text{lock}}^0 /2\pi=$ 25 Hz} clearly shows a dependence on the locking point $\delta_{\text{lock}}$ {spanning more than \SI{2e-15}{} in relative units,} as shown in Fig.~\ref{fig:2D-exc-frac}. {In order to find the absolute frequency shift, we also performed {d}ifferential frequency measurements as a function of the total lattice occupation {$N_\text{tot}^{_\text{2D}}$} at $\delta_{\rm lock}^0$, as shown in the inset of Fig.~\ref{fig:2D-exc-frac}. The interaction-induced shift exhibits a marked deviation from linearity already for {$N_\text{tot}>$} \SI{3e4}{}.}
{The two data sets are {simultaneously} fit by numerical integration of the spin model equation{, implemented with QuTiP}~\cite{Johansson2012}, and finding the clock laser detuning fulfilling (\ref{eq:lock}), and forced to share the same interaction parameters, while keeping independent absolute offsets.}
The resulting interaction values are $U_{eg}/h=${\SI{-15(2)}{Hz}}  and $U_{ee}/h=${\SI{12(4)}{Hz}, while the frequency offsets are \SI{0.16(4)}{Hz} and \SI{0.16(12)}{Hz} respectively, which are consistent with a single frequency offset corresponding to the absolute frequency shift for {$\delta_{\rm int}(N_\text{tot}^0,\delta_{\rm lock}^0)$} for both measurements}. {We therefore find that the interaction-induced frequency shift is canceled} for $\delta_{\text{lock}}/2\pi = $  {\SI{32(1)}{Hz}} {(or $\delta_{\text{lock}}/\Omega =0.94(3)$, close to the peak sensitivity point at about $\sim 0.8$~\cite{Dick1989})}. We also fit this data with the mean-field model including dephasing. The fitted dephasing rate $\gamma_{\rm deph}$ is consistent with zero. This can be explained as a consequence of the lower occupation number per lattice site and the reduced dimensionality of the transverse spatial degrees of freedom.

\begin{figure}[tb]
    \centering
    \includegraphics[width=0.99\linewidth]{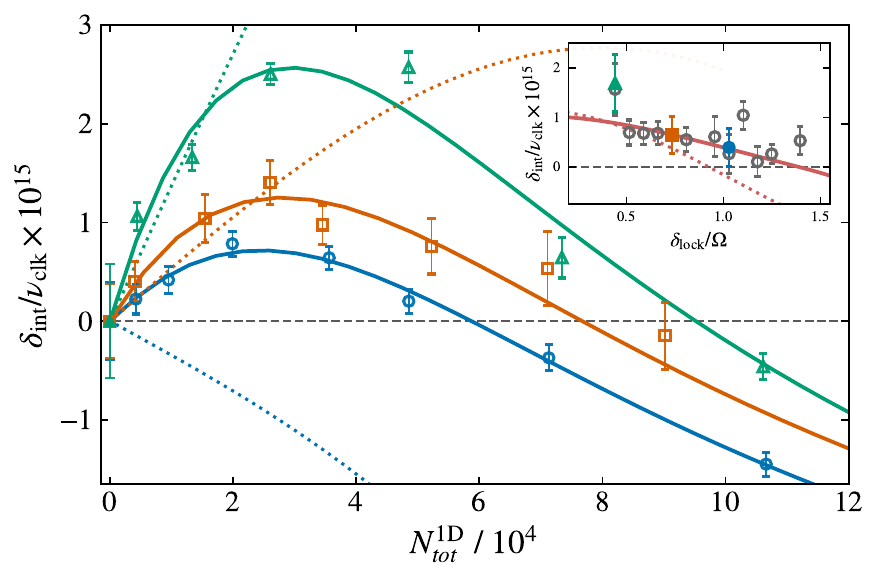}
    \caption{Measurement of the interaction frequency shift in a 1D \srb optical lattice clock. Main {panel}: interleaved {interaction} shift measurement{s} as function of the {total lattice occupation {$N_\text{tot}$}}  at {three} different lock detunings $\delta_{\text{lock}}/\Omega${: 0.4 (triangles, green), 0.7 (squares, orange), 1.0 (circles, blue)}. {Solid} lines represents numerical fit by integrating the {mean-field dissipative optical Bloch equations at the different $\delta_{\text{lock}}$ values, dashed lines are the unitary many-body spin-model interaction shift with same scattering parameters}. Inset: Interaction frequency shift at a fixed total atom number {$N_\text{tot}=7\times10^3$} as function of $\delta_{\text{lock}}$. {Colored points correspond to the frequency offsets determined from fitting the data in the main panel.} Details in the main text.}
    \label{fig:1Ddensity_shift}
\end{figure}

In the 1D lattice configuration, we measure the effect of interactions on the OLC frequency by interleaved self-comparison with respect to a clock set to {$N_\text{tot}^0=7\times10^3$} and $\delta_{\rm lock}^0/2\pi$ = \SI{25}{Hz}. The resulting differential measurement is shown in Fig.~\ref{fig:1Ddensity_shift}, where each data set corresponds to a different clock locking detuning $\delta_{\text{lock}}$.
Each data set shows at small values of {$N_\text{tot}\lesssim$} \SI{2e4}{} a positive linear interaction shift, while for higher occupation values the shift changes sign, implying a {zero crossing} at which the interaction shift can be canceled.
{Again, the} data are fit by numerical integration of the {spin model with and without including relaxation in the mean-field approximation,} and by performing a spatial averaging over all possible lattice occupation values~\cite{Rey_2014}. We find better agreement with the data for the dissipative {mean-field equations}, where the additional dephasing parameter $\gamma_{\rm deph}$ is included. However, in the 1D case, the estimated absolute value of the excited-excited two-body interaction energy $|U_{ee}|$ is larger than $|U_{eg}|$. The fit results for the {interaction energies} for the 1D lattice at 80 $E_r$ and a typical lattice temperature are {$U_{ee}/h =$ \SI{0.25(3)}{Hz} and $U_{eg}/h =$ \SI{-0.14(4)}{Hz}}, respectively; all measurements are consistent within 1$\sigma$. {The estimated dephasing rate $\gamma_{\rm deph}=$ \SI{1.3(2)}{s^{-1}}, or $\gamma_{\rm deph}/n_0=$ \SI{120(20)}{\micro\meter^{3}\,s^{-1}}, consistent with previous experiments~\cite{Lisdat2009} and an independent estimate from Rabi oscillations~\cite{suppMaterial}. This apparent discrepancy with 2D data {may be reconciled} by considering a reduced bosonic correlation coefficient $\kappa_{eg}^{(2)} = 1$ due to collapse of the $e-g$ coherence on the relevant collision timescale }(see~\cite{suppMaterial}).

We also compare the interaction-induced frequency shift at a fixed total atom number, {$N_\text{tot} = 7 \times 10^3$}, for various locking points $\delta_{\text{lock}}$, which correspond to different clock excitation fractions~\cite{Campbell2009,Lemke_2011}. These are compared with the expected shifts calculated using the interaction energies extracted from the density shift measurements, as shown in the inset of Fig.~\ref{fig:1Ddensity_shift}. The differential measurement points are corrected by the absolute frequency offset extracted by the fit performed on $N_\text{tot}$. The expected shift is comparable to the current experimental sensitivity; nevertheless, the data are consistent with the predicted trend.
We therefore find that the interaction-induced frequency shift in the 1D lattice is canceled for $\delta_{\text{lock}}/2\pi$ = \SI{48(1)}{Hz} (or $\delta_{\text{lock}}/\Omega =$ 1.4, which is {not suitable} for optimal clock operation).

To highlight the crucial role of dissipation, we also show (dashed lines in Fig.~\ref{fig:1Ddensity_shift}) the prediction of unitary dynamics, using the coupling parameters from the fit. However, we double the value of $U_{eg}$, to take into account that in this case no dissipation mechanism brings $\kappa_{eg}^{(2)}\to 1$. We observe that the data deviate from this prediction both at large values of {$N_\text{tot}$} (main panel) and at large $\delta_{\rm lock}$ (inset), where the influence of the nonlinear interaction is lower than the linear term. This implies that the expected sign change near the half-maximum point of the Rabi fringe is not observed.


\begin{table}[b]
    \centering
    \caption{Summary of the measured interaction energies {for the 1D and 2D lattices} and the derived s-wave scattering lengths for the clock states {of} the two-level \srb bosonic system.}
    \begin{tabular}{ccc}
    \toprule
    \textrm{Quantity}& \textrm{Value} & Ref.\\
    \colrule
        $(U_{ee}-U_{gg})/h$ & 0.25(3) Hz & [This work, 1D OL]  \\
        & 12(4) Hz& [This work, 2D OL]\\
        $U_{eg}/h$ & -0.14(4) Hz& [This work, 1D OL]  \\
        &-15(2) Hz& [This work, 2D OL]\\
        $a_{gg}$ &-2.2(2) $a_0$& \cite{Stellmer2013}\\
         $|a_{ee}|$& $100(50)$ $a_0$&\cite{Traverso_2009}\\
         $a_{ee}$&{105(16)} $a_0$& [This work]\\
         $a_{eg}$&{-125(12)} $a_0$&[This work]\\
    \botrule
    \end{tabular}
    \label{tab:my_label}
\end{table}

Finally, we extract the {values for the }s-wave scattering lengths from the {experimentally} estimated interaction {energies and the thermally averaged atomic volumes, which results in the main source of uncertainty~\cite{suppMaterial}}. The calculated values are reported in Tab.~\ref{tab:my_label} {after averaging among the three different experimental methods described here}. Regarding the $e-e$ scattering length, our estimate is in agreement with previously reported values~\cite{Traverso_2009}. For the inter-{level} scattering length, {the reported value $a_{eg} =$ {\SI{-125(12)}{}} $a_0$ is the first reported estimate. This can be used as a benchmark for both \textit{ab initio} quantum chemistry calculations for the \srclock interaction potential, and for clock-line photoassociation~\cite{Borkowski2018,Betterman23}.}

\paragraph{Conclusions}
In conclusion, {we report the first extensive study {of the measurement} and control of interaction-induced frequency shifts in a bosonic optical lattice clock, both varying the total lattice population over almost two orders of magnitude ({$N_\text{tot}\sim 10^3 - 10^5$)} and the clock excitation level. In the case of the 2D lattice, with two spatial degrees of freedom tightly confined and a low average site occupation number, we observe many-body features like the emergence of an interaction sideband. {We find that a collective} spin model approximation to describe the interaction dynamics is valid even in a thermal atomic ensemble. Here cancellation of the interaction-induced shift is possible in the proximity of the optimal sensitivity point. Regarding the 1D lattice, we observe reduced inter-level interaction energy and {dynamics influenced by dephasing}, possibly due to radial-mode-changing (or lateral) collisions. However, a dissipative mean-field approach {can still reproduce} the nonlinear density shift measured for this clock transition. In this way we are able to extrapolate to the interaction shift at level of \SI{3(2)e-16}{}~\cite{SalvatierraPhD26}. This approach can be useful for bosonic clocks based on other atomic species that are used for tests of fundamental physics by isotope shift measurements~\cite{Miyake2019,Ono2022}.} 
Knowledge of the coupling constants of the $gg$, $ee$ and $eg$ channels opens the intriguing perspective of realizing an XXZ spin model~\cite{Duan_ControllingSpinExchange_2003}, provided sufficiently low temperatures are attained~\cite{Chen25}. Combining this with Rabi driving in the rotating frame of the clock laser realizes the XXZ model in a transverse field, which has been proposed as a novel method to adiabatically prepare scalable spin-squeezed states~\cite{Comparin2022}.

\nocite{Breuer_TheoryOpenQuantum_2002,Gonzalez_Ballestero_Tutorialprojectorapproach_2024,Giaccari_Higherorderadiabaticelimination_2026,Bertaina-dephasing,pethick2002bose, fetter2003quantum,Bishof_Inelasticcollisionsdensitydependent_2011,Bouganne_2017}

\begin{acknowledgments}
We thank C. D'Errico, {F. Pereira dos Santos} and T. Roscilde for their expertise and for stimulating discussions. The project 23FUN02 CoCoRICO has received funding from the European Partnership on Metrology, co-financed from the European Union’s Horizon Europe Research and Innovation Programme and by the Participating States.
\end{acknowledgments}

\bibliography{finalbib}

\clearpage
\pagebreak

\renewcommand{\theequation}{S.\arabic{equation}}
\setcounter{equation}{0}
\renewcommand{\thefigure}{S\arabic{figure}}
\setcounter{figure}{0}
\renewcommand{\thetable}{S\arabic{table}}
\setcounter{table}{0}

\begin{widetext}
\begin{center}
	\textbf{\large Supplemental Material
    for \\ Measurement and control of the interaction frequency shift in bosonic optical lattice clocks
    }
\end{center}
\end{widetext}
\appendix

\section*{Spin model of interacting bosons in an optical lattice}

In this section, we aim to derive the spin model presented in Eq.~\ref{eq:Hspin} from the general many-body theoretical framework, employing adiabatic elimination techniques for open quantum systems~\cite{Breuer_TheoryOpenQuantum_2002,Gonzalez_Ballestero_Tutorialprojectorapproach_2024,Giaccari_Higherorderadiabaticelimination_2026}. A full derivation will be presented elsewhere~\cite{Bertaina-dephasing}. 

The Hamiltonian of an ensemble of trapped and interacting bosons with two internal states is~\cite{pethick2002bose, fetter2003quantum}:

\begin{eqnarray}
\hat{H} &=& \sum_{\alpha} \int d^3r \, \hat{\psi}^\dagger_\alpha(\mathbf{r}) \left( -\frac{\hbar^2}{2M} \nabla^2 + V_{\text{ext}}(\mathbf{r}) \right) \hat{\psi}_\alpha(\mathbf{r})\nonumber\\
&+& \frac{1}{2} \sum_{\alpha \beta} \int d^3r d^3r' \, \hat{\psi}^\dagger_\alpha(\mathbf{r}) \hat{\psi}^\dagger_\beta(\mathbf{r'}) V_{\alpha\beta}(\mathbf{r} - \mathbf{r'}) \hat{\psi}_\beta(\mathbf{r'}) \hat{\psi}_\alpha(\mathbf{r}),\nonumber\\
&-&\frac{\hbar\Omega}{2}\int d^3r \,\left[ \hat{\psi}^\dagger_e(\mathbf{r})
e^{-i(\mathbf{k}\cdot\mathbf r-\omega_L t)}\hat{\psi}_g(\mathbf{r}) +\text{h.c.} \right]
\end{eqnarray}
where $\bm{r}$ indicates the translational degree of freedom,  while $\alpha, \beta$ label the two internal degrees of freedom $e, g$ (e.g. hyperfine or optical clock states).  \( V_{\alpha\beta}(\mathbf{r} - \mathbf{r'}) \) describes the interactions between bosons in internal states \( \alpha \) and \( \beta \). The factor $1/2$ in the interaction term comes from avoiding double counting of pairwise interactions between bosons at the same site. $V_{\text{ext}}$ is the external trapping potential, which we assume to be independent of the internal state at the dominant order (magic wavelength condition~\cite{Katori1999}). Finally, {the last line encapsulates} the coupling of the internal levels to the external semiclassical clock laser field, modeled by Rabi coupling {frequency} $\Omega$, laser frequency $\omega_L$ and wavevector $\mathbf{k}$.

As a starting point in the 1D configuration, we assume that longitudinal motion is frozen to a single orbital $\phi_0(z)$, while the radial plane supports modes $\varphi_\mu(\bm\rho)$, and make the expansion
\begin{equation}
\hat\psi_\alpha(\bm r)=\phi_0(z)\sum_\mu \varphi_\mu(\bm\rho)\,\hat a_{\alpha\mu}\,.
\end{equation} 
An analog expansion holds in the 2D lattice configuration, with the difference that two directions are tightly confined. In the following, we indicate by $N$ the number of atoms in the considered trap well.
The Rabi-driven multimode Hamiltonian is then
\begin{align}
\hat H&=\hat H_{\rm tr}+\hat H_{\rm Rabi}+\hat H_{\rm int},\\
\hat H_{\rm tr}&=\sum_{\mu,\alpha}\epsilon_\mu \hat a^\dagger_{\alpha\mu}\hat a_{\alpha\mu},\\
\hat H_{\rm Rabi}&=-\hbar\Omega\sum_\mu \hat S_x^\mu-\hbar\delta\sum_\mu \hat S_z^\mu,
\end{align}
with mode-resolved spins
\begin{equation}
\hat S_x^\mu=\frac{\hat a^\dagger_{e\mu}\hat a_{g\mu}+\hat a^\dagger_{g\mu}\hat a_{e\mu}}{2},\quad
\hat S_z^\mu=\frac{\hat N_e^\mu-\hat N_g^\mu}{2},
\end{equation}
and $\hat N_\alpha^\mu=\hat a^\dagger_{\alpha\mu}\hat a_{\alpha\mu}$. The interaction is
\begin{equation}
\hat H_{\rm int}=
\frac12\sum_{\mu\nu\eta\lambda}\sum_{\alpha,\beta}
U_{\alpha\beta}^{\mu\nu\eta\lambda}
\hat a^\dagger_{\alpha\mu}\hat a^\dagger_{\beta\nu}\hat a_{\beta\eta}\hat a_{\alpha\lambda}\,,
\end{equation}
where the coefficients are radial overlap integrals multiplied by scattering amplitudes.

\paragraph*{Inelastic two-body channels.}
In addition to the Hamiltonian dynamics, inelastic collisions transfer atom pairs to internal or motional states that are not retained in the two-level model. After tracing over these product channels (that are untrapped and thus lost), the continuum master equation contains
\begin{equation}
\dot{\hat\rho}
=-\frac{i}{\hbar}[\hat H,\hat\rho]
+\int d^3r\,
\left[\mathcal D[\hat L_{ee}(\bm r)]
+\mathcal D[\hat L_{eg}(\bm r)]
\right]\hat\rho ,
\end{equation}
where the dissipation superoperator is $\mathcal{D}[\hat{L}]\bullet = \hat{L}\bullet\hat{L}^\dagger-\{\hat{L}^\dagger\hat{L},\bullet\}/2$ and the two-body loss operators are
\begin{align}
\hat L_{ee}(\bm r)
&=
\sqrt{\frac{\beta_{ee}}{2}}\,
\hat\psi_e(\bm r)\hat\psi_e(\bm r),
\nonumber\\
\hat L_{eg}(\bm r)
&=
\sqrt{\beta_{eg}}\,
\hat\psi_e(\bm r)\hat\psi_g(\bm r).
\label{eq:continuumloss}
\end{align}

\paragraph*{Assumptions for radial elimination.}
We trace over the radial sector and keep only an effective spin dynamics under:
(i) longitudinal freezing; (ii) stationary thermal radial state at temperature $T_r$; (iii) weak backaction of spin dynamics on the radial distribution; (iv) short radial correlation time compared with the spin times $\Omega^{-1}$, $|\delta|^{-1}$, $\hbar/|\chi|$; (v) Born factorization
\begin{equation}
\hat\rho(t)\approx \hat\rho_s(t)\otimes \hat\rho_r^{\rm th};
\end{equation}
(vi) diagonal radial thermal state in the radial occupation basis. Only radial-number-conserving terms
survive in the interaction and the reduced spin state is $\hat\rho_s=\Tr_r\hat\rho$.

The first-order reduction gives the thermally averaged Hamiltonian and renormalizes the loss channels already present in the continuum master equation. New dephasing terms generated by the Hermitian interaction appear only at second order in the centered radial fluctuations.

\paragraph*{First-order effective Hamiltonian.}
At first order, the effective spin Hamiltonian is
\begin{equation}
\hat H^{(1)}=\Tr_r\!\left(\hat H\,\hat\rho_r^{\rm th}\right).
\end{equation}
The Rabi term remains $-\hbar\Omega \hat S_x-\hbar\delta \hat S_z$, where
\begin{equation}
\hat S_\alpha=\sum_\mu \hat S_\alpha^\mu,\qquad
\hat S_z=\frac{\hat N_e-\hat N_g}{2},\qquad
\hat N_\alpha=\sum_\mu \hat N_\alpha^\mu.
\end{equation}
The $U^{\mu\nu}_{\alpha\beta}=U^{\mu\nu}_{\beta\alpha}=U^{\mu\nu\nu\mu}_{\alpha\beta}$ interaction coefficient matrix is dominated by diagonal terms $\nu=\mu$ in the harmonic confinement approximation~\cite{Rey_2014}, with off-diagonal terms strongly suppressed at typical operating temperature. Therefore, the radial-number-conserving interaction terms are:
\begin{align}
\hat H_{\rm int}^{\rm diag}
=\frac12\sum_{\mu}
&\left[
U_{ee}^{\mu\mu}\hat N_e^\mu(\hat N_e^\mu-1)
+
U_{gg}^{\mu\mu}\hat N_g^\mu(\hat N_g^\mu-1)\right.\nonumber\\
&+\left.2 U_{eg}^{\mu\mu}\hat N_e^\mu\hat N_g^\mu\right]\,.
\label{eq:Hdiagint}\end{align}

We assume thermal radial occupations $p_\mu=e^{-\beta_r\epsilon_\mu}/Z(\beta_r)$, with $\beta_r=(k_BT_r)^{-1}$ and $\sum_\mu p_\mu=1$. For $\mu=\nu$ we keep Gaussian coefficients,
\begin{align}
\Tr_r\!\left(\hat N_\alpha^\mu(\hat N_\alpha^\mu-1)\hat\rho\right)&=
\kappa_{\alpha\alpha,\mu}^{(2)}\,p_\mu^2 \hat N_\alpha(\hat N_\alpha-1)\hat\rho_s,\;\; \alpha=g,e \nonumber\\
\Tr_r\!\left(\hat N_e^\mu \hat N_g^\mu \hat\rho\right)&=
\kappa_{eg,\mu}^{(2)}\,p_\mu^2 \hat N_e\hat N_g \hat\rho_s .
\end{align}
For completely indistinguishable ideal thermal bosonic radial modes, the bunching factor is $g^{(2)}=2$ and \begin{equation}
\kappa_{ee,\mu}^{(2)}= \kappa_{gg,\mu}^{(2)}=\kappa_{eg,\mu}^{(2)}= g^{(2)},
\end{equation}
for all combinations and modes. We argue that this is the consistent choice for an initially spin-coherent state with coherent Rabi dynamics faster than dephasing and spin-dependent thermal processes. However, since Born factorization is an approximation, and we cannot completely exclude spin-dependent motional decorrelation or independent rethermalization, we have admitted that the thermal bunching factor can be different for the various spin combinations and we have tested the case $\kappa_{eg,\mu}^{(2)}\simeq 1$ which is the expected result for an incoherent $e/g$ mixture.

This yields the effective spin Hamiltonian
\begin{equation}
\hat H_{\rm int}^{(1)}=
\frac12 \bar{U}_{ee}\hat N_e(\hat N_e-1)
+\frac12 \bar{U}_{gg}\hat N_g(\hat N_g-1)
+\bar{U}_{eg}\hat N_e\hat N_g,\label{eq:Hintspin}
\end{equation}
with
\begin{equation}
\bar{U}_{\alpha\beta}=\sum_\mu U_{\alpha\beta}^{\mu\mu}\kappa_{\alpha\beta,\mu}^{(2)}\,p_\mu^2 \label{eq:SM1}
\end{equation}
We apply a Gaussian approximation to Eq.~\eqref{eq:SM1}, and get
\begin{equation}
    \bar{U}_{\alpha\beta} \simeq\kappa_{\alpha\beta}^{(2)}(4\pi\hbar^2a_{\alpha\beta})/M\,\int|w(\mathbf{r})|^4d^3r
\end{equation}
that we denote as $U_{\alpha\beta}$ in the main text, where $w(\mathbf{r})$ is a temperature-dependent spatial wavefunction approximated by Gaussian harmonic oscillator orbital functions~\cite{Swallows_2012}, whose $1/e$ widths are
\begin{equation}
    \label{eq:L}  
L_j = \sqrt{\frac{\hbar}{2\pi M \nu_j}}\times\sqrt{2\langle n_j\rangle+1} = \ell_0^j \sqrt{2\langle n_j\rangle+1}
\end{equation}
where $\nu_j$ is the trap oscillation frequency along $\hat{j}$, and
\begin{equation}
\langle n_j\rangle =\left(\exp\left[\beta_r{h\nu_j}\right]-1\right)^{-1}\,. 
\end{equation}
Writing $\hat N_e=N/2+\hat S_z$, $\hat N_g=N/2-\hat S_z$, and $C = (U_{ee}-U_{gg})/2$, $\chi = (U_{ee}+U_{gg}-2U_{eg})/2$, we obtain the effective spin Hamiltonian in the Main text, Eq.~\eqref{eq:Hspin}.

\paragraph*{Local-spin embedding and second-order dephasing.}
The strict two-mode bosonic model spans only the symmetric $J=N/2$ sector.
To expose the local structure of the reduced dissipator, we embed the
internal dynamics in the Hilbert space of $N$ spin-$1/2$ particles,
\begin{equation}
\hat S_\alpha=\sum_{i=1}^N\hat s_\alpha^{(i)},
\quad
\hat s_\alpha^{(i)}=\frac{\hat\sigma_\alpha^{(i)}}{2},
\quad
\hat P_{e,g}^{(i)}
=
\frac{\mathbbm 1}{2}\pm\hat s_z^{(i)},
\end{equation}
where the particle labels are a bookkeeping device for the symmetrized bosonic many-body state, and we also consider the radial projector $|\mu_i\rangle\langle\mu_i|$. 
It is now useful to identify the mode-resolved occupation operator with 
\begin{equation}
\hat N_\alpha^\mu =
\sum_i\hat P_\alpha^{(i)} |\mu_i\rangle\langle\mu_i|\,,
\label{eq:NfromPQ}
\end{equation}
and the radial interaction operator for an ordered particle pair with
\begin{equation}
\hat B_{ij}^{\alpha\beta} \equiv 
\sum_{\mu}
U_{\alpha\beta}^{\mu\mu}
|\mu_i\mu_j\rangle\langle\mu_i\mu_j|\,,
\end{equation}
within the same-mode approximation used in Eq.~\eqref{eq:Hdiagint}, so that the Hamiltonian can be equivalently written as  
\begin{align}
\hat H_{\rm int}^{\rm diag}
= \frac12\sum_{i\neq j} \Big[
&
\hat B_{ij}^{ee}\hat P_e^{(i)}\hat P_e^{(j)}
+\hat B_{ij}^{gg}\hat P_g^{(i)}\hat P_g^{(j)}
\nonumber\\
&+ \hat B_{ij}^{eg}
\left( \hat P_e^{(i)}\hat P_g^{(j)}
+ \hat P_g^{(i)}\hat P_e^{(j)} \right)
\Big]\,.
\label{eq:Hintpair}
\end{align}

This allows us to consider beyond-leading-order effects. In fact, since the radial thermal average of the pair interaction is $\Tr_r(\hat B_{ij}^{\alpha\beta}\,\hat{\rho}_r^{\text{th}})=\bar{U}_{\alpha\beta}$, we can write the interaction Hamiltonian~\eqref{eq:Hintpair} as the already found radially averaged part, plus fluctuations with zero thermal mean
\begin{equation}
\hat H_{\rm int}^{\rm diag} \approx \hat H_{\rm int}^{(1)} + \hat{V}_\phi
\end{equation}
up to radial fluctuation terms proportional only to the identity or total occupation operators, which do not affect the internal spin dynamics. Here,
\begin{equation}
\hat V_\phi =
\frac12\sum_{i\neq j}\sum_{\alpha=e,g}
\delta\hat\Delta_{ij,\alpha}\,
\hat P_\alpha^{(i)}\hat s_z^{(j)}
\end{equation}
and $\Tr_r(\hat{V}_\phi \hat\rho_r^{\text{th}})=0$. 
Here $\delta\hat\Delta_{ij,\alpha}=(\hat B_{ij}^{\alpha e} -\bar{U}_{\alpha e})-(\hat B_{ij}^{\alpha g}-\bar{U}_{\alpha g})$ can be interpreted as the fluctuation of the collisional
transition shift of atom $j$, conditioned on atom $i$ being in state
$\alpha$. This induces a non-unitary second-order Born-Markov contribution to the evolution of the spin density matrix~\cite{Breuer_TheoryOpenQuantum_2002,Gonzalez_Ballestero_Tutorialprojectorapproach_2024}
\begin{equation}
\dot{\hat\rho}_s\big|_\phi
= -\frac{1}{\hbar^2}\int_0^\infty d\tau\,
\Tr_r \left[ \hat V_\phi(t), \left[ \hat V_\phi(t-\tau),
\hat\rho_s(t)\otimes \hat\rho_r^{\rm th} \right] \right]\,,
\end{equation}
which, assuming short-lived and approximately pair-local radial correlations, reduces to a contribution of the Lindbladian of this form
\begin{equation}
\gamma_e\sum_{i\neq j}
\mathcal D[\hat P_e^{(i)}\hat s_z^{(j)}]\hat\rho_s
+
\gamma_g\sum_{i\neq j}
\mathcal D[\hat P_g^{(i)}\hat s_z^{(j)}]\hat\rho_s ,
\label{eq:effective_dephasing}
\end{equation}
where, schematically,
\begin{equation}
\gamma_\alpha
\sim
\frac{1}{4\hbar^2}\int_{-\infty}^{\infty}d\tau\,
\Tr_r\left[
\delta\hat\Delta_{ij,\alpha}(\tau)
\delta\hat\Delta_{ij,\alpha}(0) \hat\rho_r^{\rm th}
\right].\label{eq:gammadeph}
\end{equation}
Since the jumps in Eq.~\eqref{eq:effective_dephasing} are diagonal in the internal basis, they produce pure dephasing and do not modify the populations. This provides a microscopic motivation for a density-dependent dephasing channel of the form used phenomenologically in optical-clock models~\cite{Lisdat2009,Bishof_Inelasticcollisionsdensitydependent_2011}: the instantaneous clock shift of an atom fluctuates because its collisional shift depends on the internal state and radial state of its collision partners. 
{Following~\cite{Lisdat2009}, we considered only the term dependent on the ground-state population in Eq.~\eqref{eq:effective_dephasing}, setting $\gamma_\text{deph}=\gamma_g$ as a fit parameter in our analysis.}

\paragraph*{Two-body losses in the single-mode model.}
Under the same spatial-mode reduction used for the Hamiltonian,
Eq.~\eqref{eq:continuumloss} becomes
\begin{equation}
\hat L_{ee}
=
\sqrt{\frac{\Gamma_{ee}}{2}}\,
\hat a_e^2,
\qquad
\hat L_{eg}
=
\sqrt{\Gamma_{eg}}\,
\hat a_e\hat a_g\,,
\end{equation}
where $\Gamma_{ee},\Gamma_{eg}$ contain thermally averaged
spatial-overlap factors.

To represent the losses in a local-spin-like model, we enlarge each local
space by one vacancy state that models all possible loss states $\mathcal H_i
=\operatorname{span}\{|0\rangle_i,|g\rangle_i,|e\rangle_i\}$.
For every unordered pair $i<j$, the pair-removal jumps in this local space are thus modeled by
\begin{align}
\hat J_{ij}^{ee} &=
\sqrt{\Gamma_{ee}}\,
|0_i0_j\rangle\langle e_i e_j|,\nonumber\\
\hat J_{ij}^{eg,+} &=
\sqrt{\Gamma_{eg}}\,|0_i0_j\rangle\langle e g,+_{ij}|,
\end{align}
with $| e g,+_{ij}\rangle=(| e_i g_j\rangle+| g_i e_j\rangle)/\sqrt{2}$.
Their rate operators count the appropriate pairs:
\begin{align}
\sum_{i<j}\hat J_{ij}^{ee\dagger}\hat J_{ij}^{ee}
&=
\frac{\Gamma_{ee}}{2}\hat N_e(\hat N_e-1),
\nonumber\\
\sum_{i<j}
\hat J_{ij}^{eg,+\dagger}\hat J_{ij}^{eg,+}
&=
\Gamma_{eg}\hat N_e\hat N_g,
\end{align}
with the second equation holding exactly strictly-speaking only in symmetrized spin manifolds.

Thus the single sink state $|0\rangle$ allows the pair losses to be included
directly at the reduced spin-model level. These terms are inherited from
the original continuum Lindbladian and are not generated by the
second-order radial elimination.

\section*{Mean-field optical Bloch equations }

Let $N_0$ denote the initial number of atoms. In the
presence of loss we use the unnormalized one-body density matrix
\begin{equation}
\rho_{ee}=\frac{\langle\hat N_e\rangle}{N_0},
\quad
\rho_{gg}=\frac{\langle\hat N_g\rangle}{N_0},
\quad
\rho_{eg}=\frac{\langle\hat S_-\rangle}{N_0},
\end{equation}
whose trace
\begin{equation}
{P_\text{tot}}=\rho_{ee}+\rho_{gg}\leq 1\label{eq:ptot}
\end{equation}
is the surviving atom fraction $\langle\hat{N}\rangle/N_0$. The Bloch variables are
\begin{equation}
u=2\Re\rho_{eg},\quad
v=-2\Im\rho_{eg},\quad
w=\rho_{ee}-\rho_{gg},\label{eq:rhotoBloch}
\end{equation}
so that $\rho_{ee}=({P_\text{tot}}+w)/2$, $\rho_{gg}=({P_\text{tot}}-w)/2$.

\paragraph*{Mean-field dissipative terms.}
The dephasing channel~\eqref{eq:effective_dephasing} only affects coherences and not populations. Under permutation symmetry and mean-field factorization, it gives
\begin{equation}
\dot\rho_{eg}\big|_\phi
=
-\Gamma_\phi\,\rho_{eg},
\qquad
\dot\rho_{ee}\big|_\phi
=
\dot\rho_{gg}\big|_\phi
=
0,
\end{equation}
where
\begin{equation}
\Gamma_\phi
=
\frac{N_0-1}{2}
\left(
\gamma_e\rho_{ee}
+
\gamma_g\rho_{gg}
\right)
.
\label{eq:Gamma_phi}
\end{equation}

Conversely, the pair-removal jumps affect all components of the one-body density matrix, and give the standard two-body loss equations
\begin{align}
\dot\rho_{ee}\big|_{\rm loss}
&=
-(N_0-1)
\left(
\Gamma_{ee}\rho_{ee}
+
\Gamma_{eg}\rho_{gg}
\right)\rho_{ee},
\nonumber\\
\dot\rho_{gg}\big|_{\rm loss}
&=
-(N_0-1)\Gamma_{eg}\rho_{ee}\rho_{gg},
\nonumber\\
\dot\rho_{eg}\big|_{\rm loss}
&=
-\Gamma_{\rm loss}\rho_{eg},\label{eq:SMloss}
\end{align}
with
\begin{equation}
\Gamma_{\rm loss}
=
\frac{N_0-1}{2}
\left[
\Gamma_{eg}(\rho_{ee}+\rho_{gg})
+
\Gamma_{ee}\rho_{ee}
\right].
\label{eq:Gamma_loss}
\end{equation}
The full dissipation of the coherence is therefore
\begin{equation}
\dot\rho_{eg}\big|_{\rm diss}
=
-\left(\Gamma_\phi+\Gamma_{\rm loss}\right)\rho_{eg}.
\end{equation}
Equivalently, for the optical Bloch vector,
\begin{align}
\dot u\big|_{\rm diss}
&=
-\left(\Gamma_\phi+\Gamma_{\rm loss}\right)u,
\nonumber\\
\dot v\big|_{\rm diss}
&=
-\left(\Gamma_\phi+\Gamma_{\rm loss}\right)v,\nonumber\\
\dot w\big|_{\rm loss}
&=
-(N_0-1)\Gamma_{ee}\frac{({P_\text{tot}}+w)^2}{4},\label{eq:dissipationBloch}
\end{align}
together with Eqs.~\eqref{eq:ptot},\eqref{eq:rhotoBloch}.

\paragraph*{Inclusion of mean-field unitary terms.}
The Bloch components evolve, under the spin Hamiltonian in the Main text, Eq.~\eqref{eq:Hspin}, as
\begin{align}
\dot u&=(\delta-C(N_0-1)/\hbar)v-\frac{2\chi}{N_0\hbar}\left\langle \hat S_z\hat S_y+\hat S_y\hat S_z\right\rangle,\nonumber\\
\dot v&=-(\delta-C(N_0-1)/\hbar)u+\Omega w+\frac{2\chi}{N_0\hbar}\left\langle \hat S_z\hat S_x+\hat S_x\hat S_z\right\rangle,\nonumber\\
\dot w&=-\Omega v.
\end{align}
{Technically speaking, having introduced losses, the $N$ multiplying the linear shift $C$ should have been  promoted to the operator $\hat{N}=\hat{N}_e+\hat{N}_g$, but we set it equal to its initial value, due to the small role of losses.}
Under the mean-field closure
\begin{equation}
\langle \hat S_\alpha \hat S_\beta+\hat S_\beta \hat S_\alpha\rangle
\approx 2\langle \hat S_\alpha\rangle\langle \hat S_\beta\rangle,
\end{equation}
and including the dissipative terms~\eqref{eq:dissipationBloch}, one gets the mean-field optical Bloch equations

\begin{align}
\dot u&=(\delta-\delta\nu_{\rm int}(w))v -\left(\Gamma_\phi+\Gamma_{\rm loss}\right)u\nonumber\\
\dot v&=-(\delta-\delta\nu_{\rm int}(w))u+\Omega w -\left(\Gamma_\phi+\Gamma_{\rm loss}\right)v\nonumber\\
\dot w&=-\Omega v -(N_0-1)\Gamma_{ee}\frac{({P_\text{tot}}+w)^2}{4}\,,\label{eq:SM2}
\end{align}
where $\hbar\delta\nu_{\rm int} = C(N_0-1)+\chi N_0 w${, or $\hbar\delta\nu_{\rm int} = C(N_0 {P_\text{tot}} -1)+\chi N_0 w$, in case a loss-weighted shift is considered at the mean-field level}. {For a clock interrogation time $T_\pi \ll (N_0\Gamma_{ee})^{-1}$ the two-body loss terms in \eqref{eq:SM2} can be neglected, and the first definition of $\delta\nu_{\rm int}$ can be used, reducing our model to Eqs.~\eqref{eq:mfeqs} in the Main text.}

\section*{Relaxation rate measurements}

\begin{figure}
    \centering
    \includegraphics[width=0.95\linewidth]{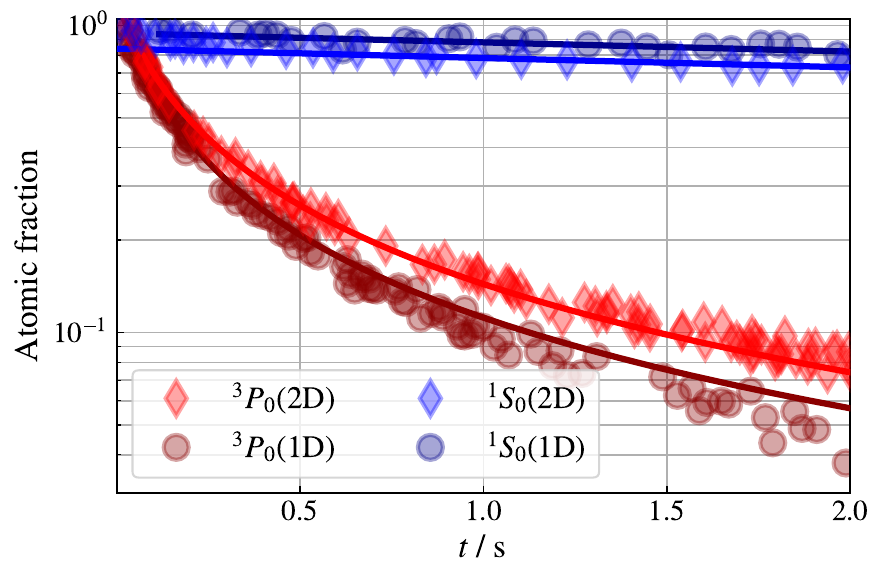}
    \caption{Atomic fraction of the ground state $P_g$ in a 1D (blue circles) and 2D (blue diamonds) and the atomic fraction of the excited state $P_e$ in a 1D (red circles) and 2D (red diamonds)  as a function of the hold time in the lattice.  }
    \label{fig:elaxation_measurements_grouped}
\end{figure}

The \srb excited clock state suffers from two-body inelastic scattering. The two-body loss rate can generally be written as

\begin{equation}
\Gamma_{\alpha\beta} = \beta_{\alpha\beta} \int d^3r \, |w(\mathbf{r})|^4 = \frac{\beta_{\alpha\beta}}{\mathcal{V}(T,U_0)},
\end{equation}
sharing the same spatial dependence on the atomic wavefunction of the elastic interactions $U_{\alpha\beta}$, condensed in the  wavefunction volume $\mathcal{V} = \mathcal{V}_0(\prod_j L^j/\ell_0^j)$. In the case of $e-e$ interaction, knowledge of the two-body density decay $\beta_{ee}$ and direct measurements of $\Gamma_{ee}$ and $U_{\alpha\beta}$ can lead to a determination of the s-wave scattering length $a_{\alpha\beta}$~\cite{Bouganne_2017}.


Assuming constant sample temperature and same site volume along the optical lattice, the time evolution of the excited clock state population for each site $N_{e,i}$ can be described as follows:
\begin{equation}
    \dot{N}_{e,i} = - N_{e,i} / \tau  - \Gamma_{ee} N_{e,i}(N_{e,i}-1)
    \label{eq:dNei}
\end{equation}
where $\tau$ is the lifetime of the atoms trapped in the lattice due to background gas scattering and lattice scattering. Integrating Eq.~\ref{eq:dNei}, and summing over all lattice sites, the time evolution of the total number of atoms $N_e(t)$, in the large number limit, can be written as follows: 
\begin{equation}
    N_{e}(t) = \sum_i N_{i,e}(0) \frac{ e^{-t / \tau }}{ 1 +   N_{i,e}(0) \tau \Gamma_{ee}   ( 1 - e^{- t / \tau})} 
    \label{eq:Ne}
\end{equation}
where $N_{i,e}(0)$ is the initial atom number in the $i$-lattice site. We assume that {$\tau$}  does not depend on the particular energy state, so that it can be measured by looking at the trap losses of the ground-state population, {which do not suffer from significant two-body losses}.

We measured the relaxation of trap populations in both the 1D and 2D lattice cases under typical experimental conditions, as shown in Fig.~\ref{fig:elaxation_measurements_grouped}. We use Eq.~\eqref{eq:Ne} as a fit function to the experimental data for both the 1D and 2D lattice geometry. In this fit procedure, the values of $\tau$ are constrained to be equal to the value extracted from $N_g$ time evolution and reported in Table~\ref{tab:fit_result_relax} for the 1D and 2D geometry. Finally, the values $N_{e,i}(0)$ for each site $i$ are computed considering a Gaussian distribution $G^{\text{1D}}_i$ ($G^{2D}_i)$ along the trapping region with a spatial dimension of $\sigma_{MOT} = \SI{192(7)}{ \micro m}$ ({$\sigma_z=\SI{70(5)}{\micro m}$} and {$\sigma_y  = \SI{12(4)}{\micro m}$)} with an initial number of atoms of $N_e(0) = \num{1.3(1)e5}$. The fit results are reported in the Tab.~\ref{tab:fit_result_relax}.

\begin{table}[h]
\begin{ruledtabular}
\begin{tabular}{lcl}
\textrm{Parameter}&
\textrm{Value}&
\textrm{Reference} \\
\colrule \\
$\tau^{\text{1D}}$ & $\SI{14(1)}{s}$ & This work \\
$\tau^{2D}$ & $\SI{14.2(4)}{s}$ & This work \\ \\
$\beta_{ee}^{\text{1D}}$ & {$\SI{21(7)}{ \micro m^3 / s}$} & This work \\ 
$\beta_{ee}^{2D}$ & {$\SI{10.0(15)}{ \micro m^3 / s}$} & This work \\ \\
$\beta_{ee}$ & $\SI{4.0(2.5)}{ \micro m^3 / s}$ & \cite{Lisdat2009} \\
 & $\SI{19(12)}{ \micro m^3 / s}$ & \cite{Traverso_2009} \\
 & $\SI{26.2(6)}{ \micro m^3 / s}$ & \cite{Dolde2025} \\
\end{tabular}
\end{ruledtabular}
\caption{\label{tab:fit_result_relax}%
Parameters and fit results obtained from the experimental data reported in Fig.~\ref{fig:elaxation_measurements_grouped}.}
\end{table}

\begin{figure}[b]
    \centering    \includegraphics[width=0.99\linewidth]{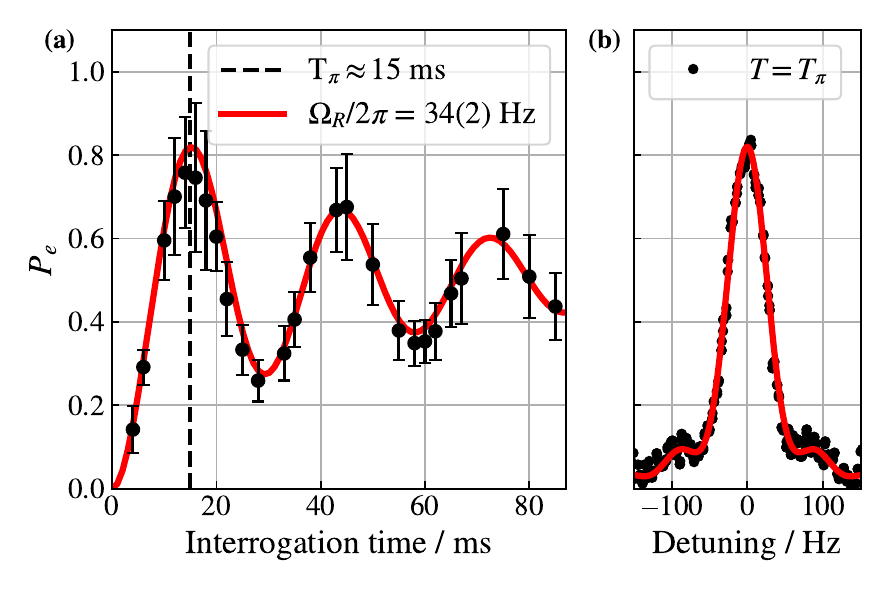}
    \caption{(a) Rabi oscillation measurement at $U_0 = 80(1)\,E_r$, and $N_\text{tot} \simeq 7\times 10^{3}$. (b) Lineshape of the clock transition at the peak of the excitation, corresponding to the Rabi $\pi$-pulse at 15 ms.}
    \label{fig:RabiOsc}
\end{figure}

We also characterize the elastic dephasing rate, introduced in the previous section, in our \srb clock by measuring Rabi oscillations at different atom numbers while keeping fixed the lattice depth. A sample measurement is shown in Fig.~\ref{fig:RabiOsc}.

For each lattice site occupation, the dynamics defined by the set of ordinary differential equations arising from \ref{eq:SM2}, including two-body losses and assuming $\delta=\delta\nu_{\rm int}$, is numerically integrated up to the interrogation time, and the total observed populations are computed by summing over all the lattice sites. In the fitting procedure, all the parameters are fixed, except the elastic dephasing coefficient $\gamma_{\rm deph}$ and the effective Rabi frequency $\Omega$. The resulting density dephasing rate is $\gamma_{\rm deph}\mathcal{V}=$\SI{186(16)}{\micro\meter^{3}\,s^{-1}}. This corresponds to a dephasing rate at typical lattice clock conditions of about \SI{1.3}{s^{-1}}.


\end{document}